\RequirePackage{lineno}
\documentclass[twocolumn,aps,superscriptaddress,showpacs,nofootinbib,floatfix]{revtex4}
\usepackage{listings}
\usepackage[utf8]{inputenc}
\usepackage{hyperref}
\usepackage{amsmath}
\usepackage{amssymb}
\usepackage{amsfonts}
\usepackage{tabularx}
\usepackage{booktabs}
\usepackage{graphicx}
\usepackage{color}

\usepackage{multirow}
\usepackage[inline]{enumitem}
\usepackage[T2A,T1]{fontenc}
\graphicspath{{fig/}}
\definecolor{theblue}{RGB}{0,50,230}

\usepackage{hyperref}
\hypersetup{
  colorlinks=true,
  linkcolor=theblue,
  citecolor=theblue,
  urlcolor=theblue
}

\newcommand {\avg}[1]{\ensuremath{\langle\kern-1.0pt\langle#1\rangle\kern-1.0pt\rangle}}

\def\eq{{\,=\,}}

\newlength\cmsFigWidth

\def\pt{p_T}
\def\bq{\begin{eqnarray}}
\def\eq{\end{eqnarray}}

\usepackage[normalem]{ulem}  

\renewcommand\sout{\bgroup \color{red} \ULdepth=-.5ex \ULset}

\begin{document}

\title{Centrality and transverse momentum dependence of hadrons in Pb+Pb collisions at LHC}


\author{Lilin Zhu}\email{zhulilin@scu.edu.cn}
\affiliation{Department of Physics, Sichuan University, Chengdu 610064, China}
\author{Hua Zheng}\email{zhengh@snnu.edu.cn}
\affiliation{School of Physics and Information Technology, Shaanxi Normal University, Xi'an 710119, China}
\author{Rudolph C. Hwa}\email{hwa@uoregon.edu}
\affiliation{Institute of Fundamental Science, University of Oregon, Eugene, OR 97403-5203, USA}

\begin{abstract}
The transverse momentum spectra of seven identified hadrons produced in Pb+Pb collisions at $\sqrt{s_{NN}}=2.76$ and 5.02 TeV at the CERN Large Hadron Collider (LHC) have been investigated in the framework of the recombination model (RM). Soft, semihard and hard partons all play important roles in our model and are uniformly treated for all hadrons produced.The investigation has been extended to non-central collisions and the parameters controlling the momentum degradation of semihard partons have been tuned. Our study shows good agreement between theoretical results and experimental data. To be able to provide a coherent explanation for the production of all identified hadrons ($\pi$, p, K, $\Lambda$, $\phi$, $\Xi$, $\Omega$) for transverse momenta as high as 14 GeV/c (for $\pi$, p and K), and for all centralities, is an unprecedented achievement that supports the RM as a sensible model to combine various mechanisms of parton production with a universal scheme of hadronization. 

\end{abstract}

\pacs{25.75.Nq, 25.75.Ld}
\keywords{}
\maketitle

\section{introduction}
The ultimate goal of the physics program with ultrarelativistic nucleus-nucleus collisions is to understand the properties of strongly interacting matter under extreme conditions of high temperature and density. Quantum Chromodynamics (QCD) predicts that at sufficiently high density strongly interacting matter undergoes a phase transition from a state of hadronic constituents to quark and gluon plasma (QGP). One of the most promising signals of the deconfinement is related to particular properties of transverse momentum spectra of final hadrons. The conventional methods to describe hadron production in heavy-ion collisions are by use of hydrodynamical model for transverse momentum $p_T<2$ GeV/c \cite{kh, tat, gjs, hydroepjczhao} and by jet fragmentation for $p_T>8$ GeV/c \cite{pq2, mt, jet}. In the intermediate region neither approaches are applicable, while parton recombination or coalescence (ReCo) model in heavy-ion collisions has been found to be more relevant \cite{hy0, gkl, fmnb}. The large baryon-to-meson ratio is an observed phenomenon that is successfully explained by ReCo to fill the gap \cite{hy0, gkl, fmnb, hz2, Zhu:2014csa}. 

Despite differences in detail, the basic ideas in the three formulations of ReCo are very similar \cite{hy0, gkl, fmnb}. In this study, the Hwa-Yang recombination model (RM) is adopted \cite{hy0}. It is formulated in one dimension along the direction of recombination on the basis that non-collinear partons have low probability of coalescence, but it incorporates fragmentation as a component of recombination so that there is a smooth transition from low to high $p_T$. Within the framework of RM, the transverse momentum and centrality dependence of hadron production in Au+Au collision at $\sqrt{s_{NN}}=200$ GeV at RHIC were well reproduced \cite{hz2}. It is found that the recombination of thermal and shower partons is the major component for intermediate $p_T$ region. In heavy-ion collisions above 2 TeV, the density of minijets produced by semihard scatterings of partons can be so high that the conventional  treatment of such collisions may be inadequate. Recently, the spectra of hadrons produced in central Pb+Pb collisions at $\sqrt{s_{NN}}=2.76$ TeV has been studied with $p_T$ up to 20 GeV/c by the recombination model. It is found that the minijets are abundantly produced at LHC compared with RHIC \cite{Zhu:2014csa}. These previous studies \cite{hz2, Zhu:2014csa} have revealed that the formation of hadrons at intermediate $p_T$ region is sensitive to the momentum degradation of the hard and semihard partons. Furthermore, due to the geometrical configuration of colliding system, the momentum degradation from the initial parton momentum to the final momentum at the medium surface is different for Au+Au and Pb+Pb collisions. Even for the same colliding system, the momentum degradation is not the same for different colliding energies. We will give more discussion later.

In heavy-ion collisions there are various theoretical issues related to minijets that have not yet evolved to a mature subject with general acceptance. The medium effects on semihard partons are important but hard to make precise, and the hadronization process is still controversial. The shower partons not only depend on the momentum degradation in the medium, but also can be hadronized through various channels, such as recombination with thermal partons on the one hand and with other shower partons on the other. At LHC the high density of jets enables the possibility of shower partons from different jets overlapping in common spatial proximity so that the contribution from their coalescence cannot be ignored. We have studied the minijet contribution to the central Pb+Pb collisions at $\sqrt{s_{NN}}=2.76$ TeV and reproduce the hadron production very well \cite{Zhu:2014csa}. Therefore, we will extend the investigation of the $p_T$ spectra of identified hadrons to non-central collisions in Pb+Pb collisions at both $\sqrt{s_{NN}}=2.76$ and 5.02 TeV with the recombination model in this study and conclude with a summary on our general view of the hadronization process in nuclear collisions.

The paper is organized as follows: In Section~\ref{RM}, we briefly review the basic framework of Hwa-Yang recombination model, including the inclusive distribution of minijets. The formalisms of parton combination for pion, proton and kaon are shown  in Section~\ref{IDH}. In Section~\ref{results}, we show results from our study on the centrality dependence of transverse momentum spectra of seven identified hadrons in Pb+Pb collisions at $\sqrt{s_{NN}}=2.76$ and 5.02 TeV. Finally, section~\ref{summary} summarizes the results and gives the conclusion from the present study.

\section{basic framework of recombination model}\label{RM}

Let us start by recalling the basic elements of the recombination model. Our study here is limited to the midrapidity and all the formulae are averaged over the azimuthal angles. Therefore, the invariant distributions of mesons and baryons are, which are averaged over $\eta$ at midrapidity \cite{hy0, hy1, hz1, hz2, Hwa:2009tx, Zhu:2014csa}
\begin{eqnarray}
\hspace{-0.5cm}p^0{dN^M\over dp_T}&=&\int {dp_1\over p_1}{dp_2\over p_2} F_{q_1\bar q_2}(p_1,p_2) R_{q_1\bar q_2}^M(p_1,p_2,p_T), 
\label{2.1}
\end{eqnarray}
\begin{eqnarray}
\hspace{-1.5cm}p^0{dN^B\over dp_T}&=&\int {dp_1\over p_1}{dp_2\over p_2}{dp_3\over p_3} F_{q_1q_2q_3}(p_1,p_2,p_3)\nonumber\\ 
&&\times {R}_{q_1q_2q_3}^B(p_1,p_2,p_3,p_T),   
\label{2.2}
\end{eqnarray}
with the transverse momenta of coalescing quarks $p_i$. $R^{M}$ and $R^{B}$ are the recombination functions (RFs) of the corresponding quarks for mesons and baryons, respectively. The parton distributions can be partitioned into various components, represented symbolically by
\begin{eqnarray}
F_{q_1\bar q_2}=\mathcal T\mathcal T+\mathcal T\mathcal S+\mathcal S\mathcal S,
\label{2.3}
\end{eqnarray}
\begin{eqnarray}
F_{q_1 q_2 q_3}=\mathcal T\mathcal T\mathcal T+\mathcal T\mathcal T\mathcal S+\mathcal T\mathcal S\mathcal S+\mathcal S\mathcal S\mathcal S,
\label{2.4}
\end{eqnarray}
where $\mathcal{T}$ and $\mathcal{S}$ represent the invariant distributions for thermal and shower partons just before hadronization, respectively.
The former contains the medium effect, while the latter is due to semihard and hard scattered partons. The consideration of shower partons is a unique feature of our model to recombination, which is empowered by the possibility to include fragmentation process as $\mathcal S\mathcal S^{1j}$ or $\mathcal S\mathcal S\mathcal S^{1j}$ recombination.
For the contribution with two or three shower partons, we need to take into account their sources. For a visualization of the various processes, we refer to the schematic diagrams in Ref. \cite{Zhu:2014csa}.  The shower partons in $\mathcal S\mathcal S$ and $\mathcal T\mathcal S\mathcal S$ can be from the same jet or two jets. For $\mathcal S\mathcal S\mathcal S$, only the contributions of $\mathcal S\mathcal S\mathcal S^{1j}$ and $\mathcal S\mathcal S\mathcal S^{2j}$ are included in this study, because the contribution from $\mathcal S\mathcal S\mathcal S^{3j}$ referring to three shower partons from three jets is highly suppressed.

The thermal parton distribution is assumed to have a simple exponential form \cite{hy0, Zhu:2014csa}
\begin{eqnarray}
\mathcal T_j(p_i)=p_i\frac{dN_j}{dp_i}=C_jp_ie^{-p_i/T_j},
\label{2.5}
\end{eqnarray}
where the subscript $j$ denotes quark ($u, d, s$). Since the mass of $s$ is different from $(u, d)$, to distinguish them, we use $q$ to denote $(u, d)$. The normalization factor $C_j$ has the dimension of inverse momentum. 

We remark with emphasis that the symbols ($\cal T$, $T_j$) and word (thermal) used here are carried over from previous works \cite{hy0, Zhu:2014csa} without modification in order to preserve the continuity of the mathematical formalism of the RM. However, the physical content of the model before hadronization may evolve with increasing energy and with better understanding of the physical processes. In particular, at the collision energies that we consider here, rapid equilibration would not be realistic, since some hard and semi-hard partons can take over 5 fm/c to traverse the transverse dimensions of the initial system, whereby they deliver considerable energy to the expanding system. Even though the bulk system may not be fully equilibrated at the time of hadronization, we still refer to the soft sector as “thermal”, so as to distinguish it from the harder sector that includes semi-hard or hard partons that produce the shower partons. The exponential form in Eq. (\ref{2.5}) is assumed on phenomenological grounds with the value of $T_j$ to be adjusted to fit the new data. The parameter $T_j$ is referred to as inverse slope, which, we stress, is not temperature, a term that is suitable only for a fully equilibrated thermal system. What is remarkable is that the exponential form in Eq. (\ref{2.5}) provides excellent fits of the data with the inverse slope being grossly different from the conventional temperature of thermalized systems at lower collision energies.

A universal formula for the energy dependence of $T_j$ was obtained in Ref. \cite{Hwa:2018qss},
\bq
T_q(s)&=&T_1 f(s), \qquad  T_s(s)=T_2 f(s),    \label{2.6}  \\
f(s)&=&{\sqrt s}\ ^\nu,   \label{2.7}
\eq
with the parameters
\bq
T_1=0.35\ {\rm GeV/c}, \quad T_2=0.46\ {\rm GeV/c}, \quad \nu=0.105.  \label{2.8}
\eq
For simplicity, we have used $\sqrt{s}$ to replace $\sqrt{s_{NN}}$. It should be pointed out that $\sqrt{s}$ in Eq. (\ref{2.7}) is in units of TeV and should therefore be regarded as a dimensionless number. The values of $T_j$ for $\sqrt{s}=2.76$ and 5.02 TeV are given in Table \ref{tab1}. 

\begin{table}
\tabcolsep0.25in
\begin{tabular}{ccc}
\hline
\hline
 $\sqrt{s}$ (TeV) &2.76 &5.02\\ 
 \hline
$T_q$ (GeV) & 0.39 & 0.415 \\
$T_s$ (GeV) &0.51 &0.545 \\
 \hline
  \hline
 \end{tabular}
 \caption{Parameters $T_q$ and $T_s$ for Pb+Pb collisions at $\sqrt{s}=2.76$ and 5.02 TeV, respectively.} 
 \label{tab1}
 \end{table}
 
As discussed in Ref. \cite{Zhu:2014csa}, the shower parton distribution is given by
\begin{eqnarray}
{\cal S}^j(p_2)=\int {dq\over q}\sum_i \hat F_i(q) S_i^j(p_2, q),  \label{2.9}
\end{eqnarray}
 where  $S_i^j(p_2, q)$ is the shower parton distribution (SPD) in a jet of type $i$ fragmentating into a parton of type $j$ with momentum fraction of $p_2/q$. It is determined by the fragmentation functions (FFs) known from analyzing leptonic processes on the basis that hadrons in a jet are formed by recombination of the shower partons in the jet \cite{hy4}. Once the SPDs are known, one can give a more complete description of hadronization, especially for the nuclear collisions. After including the centrality dependence, Eq. (\ref{2.9}) should be modified as
\begin{eqnarray}
{\cal S}^j(p_2, c)=\int {dq\over q}\sum_i \hat F_i(q, c) S_i^j(p_2, q).  \label{3.6}
\end{eqnarray}
$\hat{F_i}(q, c)$ is the distribution of hard and semihard parton of type $i$ at the medium surface after momentum degradation while traversing the medium. At a specified centrality $c$ (e.g. $c=0.05$ stands for $0-10\%$ centrality) it is defined as 
 \begin{eqnarray}
\hat F_i(q, c)&=&\frac{1}{2\pi}\int d\phi\int d\xi P_i(\xi, \phi, c) \\ \nonumber
&&\times\int dkkf_i(k, c)G(k, q, \xi), \label{2.10}
\end{eqnarray}
$P_i(\xi, \phi, c)$ is the probability for parton $i$ having a dynamical path length $\xi$ at angle $\phi$ initiated at position $(x_0, y_0)$, weighted by the nuclear overlap function and integrated over all $(x_0, y_0)$. The connection between geometry and dynamics is imbedded in $P_i(\xi, \phi, c)$. The dynamical path length $\xi$ is proportional to the geometrical path length $l$. Therefore, $P_i(\xi, \phi, c)$ can be written as
\begin{eqnarray}
P_i(\xi, \phi, c) =\int dx_0dy_0Q(x_0, y_0, c)\delta(\xi-\gamma_i l).
\end{eqnarray}
where $Q(x_0, y_0, c)$ is the probability that a hard (or semihard) parton is produced at $(x_0, y_0)$, which can be calculated from nucleon thickness functions \cite{hz2, Hwa:2009tx}. The factor $\gamma_i$ is introduced to account for the effects of jet quenching in the medium that results in additional parton degradation due to the soft partons created. The dependence of that factor on the colliding energy will be considered below. As stated in Ref. \cite{Hwa:2009tx}, all physical lengths are in units of the effective nuclear radius $R_A$, so the geometrical path length $l$ is numerical in that unit, whereas $\gamma_i$ has the characteristic of inverse length, resulting in $\xi$ being dimensionless, although it has been referred to as dynamical path length.

The function $f_i(k, c)$ is the density of parton $i$ in the phase space $kdk$. For central collisions, the initial momentum distribution $f_i(k, c)$ was parametrized as \cite{sgf}
\begin{eqnarray}
f_i(k, 0.05)=K\frac{A}{(1+k/B)^{n}}. 
\label{2.12}
\end{eqnarray}
For Pb+Pb collisions at 2.76 TeV, the parameters in Eq. (\ref{2.12}) were obtained by logarithmic interpolations of the parameters $\ln A$, $B$ and $n$ between Au+Au collisions at 200 GeV and Pb+Pb collisions at 5.5 TeV with $K=2.5$, as shown in Ref. \cite{Zhu:2014csa}. Following the same approach, we can get the parameters for central Pb+Pb collisions at 5.02 TeV listed in TABLE \ref{tab2}. Taking into account the centrality dependence, the minijet distribution can be calculated by
\begin{eqnarray}
f_i(k, c) = \frac{T_{AA}(c)}{T_{AA}(0.05)}f_i(k, 0.05).
\end{eqnarray}
The nuclear thickness function $T_{AA}(c)$ for Pb+Pb collisions are available in Ref. \cite{Abelev:2013qoq}.

For the momentum degradation function $G(k, q, \xi)$ due to energy loss, as discussed in Ref. \cite{Zhu:2014csa}, we take the simple form 
\begin{eqnarray}
G(k, q, \xi)=q\delta(q-ke^{-\xi}) \label{2.11}
\end{eqnarray}
 as an adequate approximation of the complicated processes involved in the parton medium interaction. 
 
Therefore, $\hat{F}_i(q, c)$ can be calculated once $\gamma_i$ is known. Unfortunately, we cannot calculate $\gamma_i$ directly, since it is a factor that approximates the effects of energy loss during the passage of the parton through the non-uniform and expanding medium that is not thermalized due to the energy degradation processes of all the semi-hard partons produced throughout the medium. Those effects have not been successfully treated from first principles. We have used $\gamma_i$ as a phenomenological factor with parameters adjusted to fit the data. Since quarks and gluons lose their energies at different rates when they go  through the medium, we assumed that $\gamma_q=\gamma_g/2=0.07$ and obtained excellent fits for hadron spectra in Au+Au collisions at $\sqrt{s_{NN}}=200$ GeV  \cite{hz2} . For the momentum degradation at LHC, the $\gamma_i$ factor cannot remain the same as that at RHIC due to many more minijets produced and the different geometrical configuration of colliding system. For Pb+Pb collisions, we parametrized it as \cite{Zhu:2014csa}
\begin{eqnarray}
\gamma_g(q)=\frac{\gamma_0}{1+(q/q_0)^2}.
\end{eqnarray}
The parameters $\gamma_0$ and $q_0$ are determined by fitting the spectra in the intermediate $p_T$ region. For Pb+Pb collisions at 2.76 TeV, we have obtained $\gamma_0=2.8$ and $q_0=7$ GeV/c \cite{Zhu:2014csa}. Their values for  Pb+Pb collisions at 5.02 TeV will be discussed in Sec. \ref{results}.

\begin{table*}
\tabcolsep0.2in
\begin{tabular}{|c|c|c|c|c|c|c|}
\hline
 & $g$ &  $u$ &  $d$ & $\bar u$ &  $\bar d$  &s, $\bar s$\\ 
 \hline
 $A$ [$10^4$/GeV$^2$] & 11.2 &2.02 &2.28 &0.42 &0.40& 0.154\\
 $B$ [GeV]& 0.80 &0.59 &0.58 &0.75 &0.76 &0.93\\
 $n$ & 5.68 &5.31 &5.29 &5.52 &5.53 &5.63\\
 \hline
 \end{tabular}
 \caption{Parameters for $f_i(k, 0.05)$ in Eq.\ (\ref{2.12}) for central Pb+Pb collisions at $\sqrt{s}=5.02$ TeV.} 
 \label{tab2}
 \end{table*}

\section{Inclusive distributions of hadrons}\label{IDH}
After the semihard parton distributions $\hat{F}(q ,c)$ for all species and all centralities are obtained, we can be more explicit about hadron formation by recombination. The formulae for recombination of thermal and shower partons at central collisions have been developed previously \cite{Zhu:2014csa}. We generalize them to non-central collisions here for pion, kaon, and proton production. The centrality dependence of other mesons and hyperons production can be derived in similar way. 
\subsection{Pion production}
The RF for pion is given in Refs. \cite{hy0, hy1, hz1, hz2, Zhu:2014csa}
\begin{eqnarray}
R^{\pi}(p_1, p_2, p_T)=\frac{p_1p_2}{p_T}\delta(p_1+p_2-p_T).  \label{3.1}
\end{eqnarray}
Following Eqs. (\ref{2.1}), (\ref{2.2}) and (\ref{3.1}), the four components for pion production are 
\begin{eqnarray}
{dN^{TT}_{\pi}\over p_Tdp_T} &=&[1+u(p_T, N_{part})]\frac{C^2}{6}e^{-p_T/T} ,   \label{3.2}\\
{dN_{\pi}^{TS}\over p_Tdp_T} &=& {C\over p_T^3} \int_0^{p_T} dp_1 p_1e^{-p_1/T} \\ \nonumber
&&\times\left[{\cal S}^{u}(p_T-p_1, c) +{\cal S}^{\bar d}(p_T-p_1, c)\right] ,    \label{3.3} \\
{dN^{{SS}^{1j}}_{\pi}\over p_Tdp_T} &=& {1\over p_T} \int {dq\over q^2} \sum_i \hat{F}_i(q, c)D^{\pi}_i(p_T,q) ,   \label{3.4}\\
{dN_{\pi}^{{SS}^{2j}}\over p_Tdp_T} &=& {\Gamma\over p_T^3} \int_0^{\pt} dp_1  {\cal S}^{u}(p_1, c) {\cal S}^{\bar d}(p_T-p_1, c) .     \label{3.5} 
\end{eqnarray}
The extra factor $u(p_T, N_{part})$ in Eq. (\ref{3.2}) is to describe the contribution from the resonance decays, which dominates the pion distribution at $p_T<2$ GeV/c. It will be specified below. The shower-shower recombination from one jet (${SS}^{1j}$) is equivalent to fragmentation, so we can use the FFs $D^{\pi}_i$ directly in Eq. (\ref{3.4}). $\Gamma$ in Eq. (\ref{3.5}) is the probability that two parallel partons originated from two jets can recombine. As done in Ref. \cite{Zhu:2014csa}, $\Gamma$ was estimated as the ratio of pion diameter to nucleus diameter, which is about $0.1$.

\begin{figure}[pht]
\includegraphics[width=0.45\textwidth]{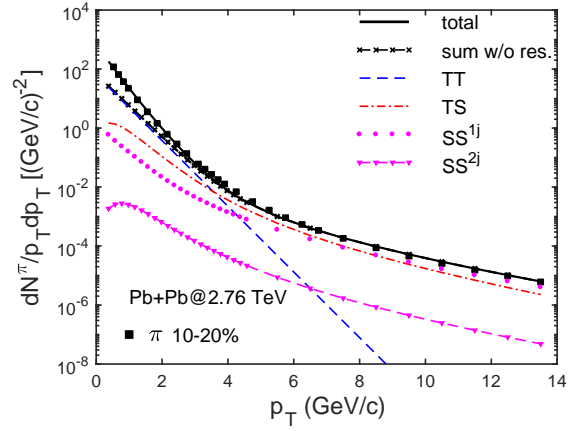}
 \caption{(Color online) Transverse momentum spectrum of pion at the centrality of 10-20\% in Pb+Pb collisions at $\sqrt{s_{NN}}=2.76$ TeV. The data are taken from Ref. \cite{Adam:2015kca}.}
 \label{fig3}
\end{figure}

\begin{figure}[pht]
\includegraphics[width=0.45\textwidth]{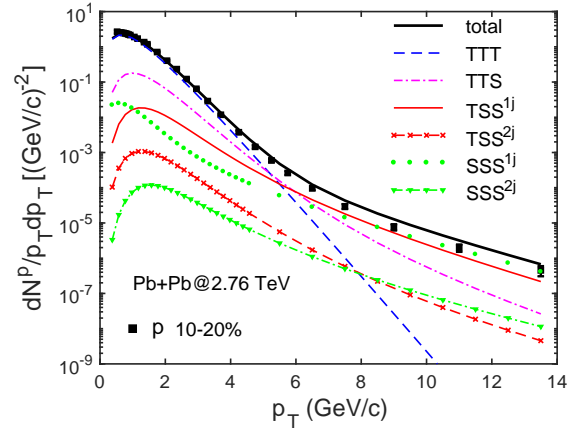}
 \caption{(Color online) Transverse momentum spectrum of proton at the centrality of 10-20\% in Pb+Pb collisions at $\sqrt{s_{NN}}=2.76$ TeV. The data are taken from Ref. \cite{Adam:2015kca}.}
 \label{fig4}
\end{figure}

\subsection{Proton production}
Proton mass is not negligible, compared to that of pion; thus, $p^0$ in Eq. (\ref{2.2}) should be replaced by the transverse mass $m_T^p=\sqrt{p_T^2+m_p^2}$. The RF for proton is given in Refs. \cite{hy0, hy1, hz1, hz2, Zhu:2014csa}, which includes the momentum conservation $\delta$ function. The thermal-thermal-thermal (TTT) recombination is 
\begin{eqnarray}
\frac{dN_p^{TTT}}{p_Tdp_T}=g_{st}^pg_pg_p'\frac{C^3p_T^2}{m_T^p}e^{-p_T/T}, \label{3.7}
\end{eqnarray}
where $g_{st}^p=1/6$ and
\begin{eqnarray}
g_p=[B(\alpha+1, \alpha+\beta+2)B(\alpha+1, \beta+1)]^{-1}, \label{3.8}
\end{eqnarray}
\begin{eqnarray}
g_p'=B(\alpha+2, \beta+2)B(\alpha+2, \alpha+\beta+4). \label{3.9}
\end{eqnarray}
$B(\alpha, \beta)$ is the beta function with $\alpha=1.75$ and $\beta=1.05$.

For thermal-thermal-shower (TTS), thermal-shower-shower (TSS) and shower-shower-shower (SSS) recombination, we have
\begin{widetext}
\begin{eqnarray}
{dN_p^{TTS}\over \pt d\pt}&=&{g_{st}^pg_p C^2\over m_T^p \pt^{2\alpha+\beta+3}} \int_0^{\pt} dp_1 \int_0^{\pt-p_1} dp_2\ e^{-(p_1+p_2)/T}  \left\{ (p_1p_2)^{\alpha+1}(\pt-p_1-p_2)^{\beta} {\cal S}^d(\pt-p_1-p_2, c)\right.\nonumber\\
&&\left.+p_1^{\alpha+1}p_2^{\beta+1}(\pt-p_1-p_2)^{\alpha} {\cal S}^u(\pt-p_1-p_2, c)\right\}, \label{3.10}
\end{eqnarray}

\begin{eqnarray}
{dN_p^{{TSS}^{1j}}\over \pt d\pt}&=&{g_{st}^pg_p C\over m_T^p \pt^{2\alpha+\beta+3}} \int_0^{\pt} dp_1 \int_0^{\pt-p_1} dp_2\ e^{-p_1/T}\left\{ p_1^{\beta+1}p_2^{\alpha}(\pt-p_1-p_2)^{\alpha} {\cal S}^{uu}(p_2,\pt-p_1-p_2, c)\right.\nonumber\\
&&\left.+p_1(p_1p_2)^{\alpha}(\pt-p_1-p_2)^{\beta} {\cal S}^{ud}(p_2,\pt-p_1-p_2, c)\right\},\label{3.11}
\end{eqnarray}

\begin{eqnarray}
{dN_p^{{TSS}^{2j}}\over \pt d\pt}&=&{g_{st}^pg_p  C\Gamma\over m_T^p \pt^{2\alpha+\beta+3}} \int_0^{\pt} dp_1 \int_0^{\pt-p_1} dp_2\ e^{-p_1/T}  \left\{ p_1^{\beta+1}p_2^{\alpha}(\pt-p_1-p_2)^{\alpha} {\cal S}^u(p_2, c) {\cal S}^{u}(\pt-p_1-p_2, c)\right.\nonumber\\
&&\left.+p_1(p_1p_2)^{\alpha}(\pt-p_1-p_2)^{\beta} {\cal S}^u(p_2, c) {\cal S}^{d}(\pt-p_1-p_2, c)\right\}, \label{3.13}
\end{eqnarray}

\begin{eqnarray}
{dN_p^{{SSS}^{1j}}\over \pt d\pt}=\frac{1}{m_T^p}\int\frac{dq}{q^2}\sum\limits_i\hat F_i(q, c)D_i^p(p_T, q),\label{3.12}
\end{eqnarray}

\begin{eqnarray}
{dN_p^{{SSS}^{2j}}\over \pt d\pt}&=&{g_{st}^pg_p\Gamma \over m_T^p \pt^{2\alpha+\beta+3}} \int_0^{\pt} dp_1 \int_0^{\pt-p_1} dp_2 \left\{ p_1^{\beta}p_2^{\alpha}(\pt-p_1-p_2)^{\alpha} {\cal S}^d(p_1, c) {\cal S}^{uu}(p_2,\pt-p_1-p_2, c)\right.\nonumber\\
&&\left.+(p_1p_2)^{\alpha}(\pt-p_1-p_2)^{\beta} {\cal S}^u(p_1, c) {\cal S}^{ud}(p_2,\pt-p_1-p_2, c)\right\}, \label{3.14}
\end{eqnarray}

where
\begin{eqnarray}
\mathcal{S}^{qq}(p_2, p_3, c)=\int\frac{dq}{q}\sum\limits_i\hat F_i(q, c){\rm S}_i^q(p_2, q){\rm S}_i^q(p_3, q-p_2).\label{3.15}
\end{eqnarray}
\end{widetext}

\subsection{Kaon production}
The four components of kaon distribution are very similar to those of pion. The differences are in the constituent quark masses between $m_q$ and $m_s$ resulting in asymmetry of the RF for kaon \cite{hwaprc2007} and the different $T_q$ of light quarks and $T_s$ of $s$ quarks. The kaon production is written as
\begin{eqnarray}
{dN^{TT}_{K}\over p_Tdp_T} &=&
{12CC_s\over m_T^Kp_T^5} \int_0^{\pt} dp_1 p_1(\pt-p_1)^2 \nonumber\\
&&\times p_1e^{-p_1/T}(p_T-p_1)e^{-(p_T-p_1)/T_s} ,     \label{3.16}
\end{eqnarray}

\begin{eqnarray}
{dN_{K}^{TS}\over p_Tdp_T} &=& {12\over m_T^Kp_T^5} \int_0^{\pt} dp_1 p_1^2(\pt-p_1)^2\nonumber  \\
&&\left[Ce^{-p_1/T}{\cal S}^{\bar s}(p_T-p_1,c)\right. \nonumber  \\
&&\left. +C_s\left({\pt\over p_1}-1\right)e^{-(\pt-p_1)/T_s}{\cal S}^u(p_1,c)\right] ,    \label{3.17} 
\end{eqnarray}

\begin{eqnarray}
{dN^{{SS}^{1j}}_{K}\over p_Tdp_T} &=& {1\over m^K_T} \int {dq\over q^2} \sum_i \hat{F}_i(q ,c)D^{K}_i(p_T,q)
,  \label{3.18}
\end{eqnarray}

\begin{eqnarray}
{dN_{K}^{{SS}^{2j}}\over p_Tdp_T} &=& {12\Gamma\over m_T^Kp_T^5} \int_0^{\pt} dp_1 p_1(\pt-p_1)^2  \nonumber  \\
&&\times{\cal S}^{u}(p_1,c) {\cal S}^{\bar s}(p_T-p_1,c) .    \label{3.19} 
\end{eqnarray}

 \begin{figure*}[t]
 \centering
\includegraphics[width=0.9\textwidth]{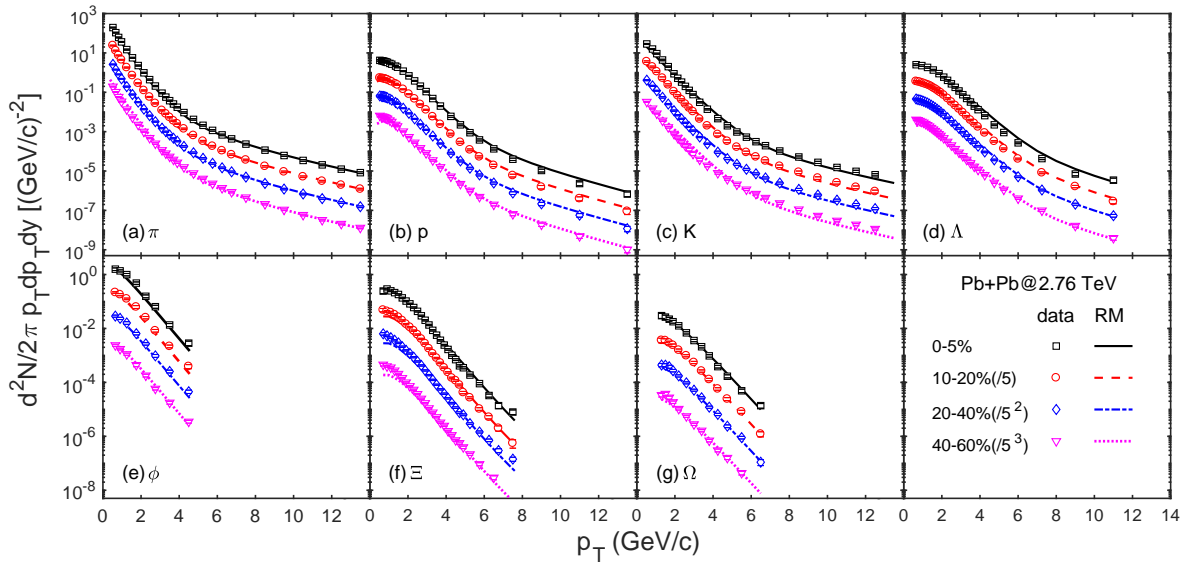}
  \caption{(Color online) Transverse momentum spectra of pion (a), proton (b), kaon (c), $\Lambda$ (d), $\phi$ (e), $\Xi$ (f) and $\Omega$ (g) from the recombination model at midrapidity in Pb+Pb collisions at $\sqrt{s_{NN}}=2.76$ TeV for different centrality classes. Scale factors are applied for better visibility. The data are taken from Ref. \cite{Adam:2015kca} for pion, kaon and proton, Ref. \cite{Abelev:2014uua} for $\phi$,  Ref. \cite{Abelev:2013xaa} for $\Lambda$ and Ref. \cite{ABELEV:2013zaa} for $\Xi$ and $\Omega$.}
\label{fig5}
\end{figure*}

\section{Results}\label{results}
So far, we have treated the momentum degradation of minijets and determined the distribution of semihard parton $\hat{F}_i(q, c)$ for any centrality and the formalism for hadronization in Secs. \ref{RM} and \ref{IDH}, respectively.  The hadron production in central Pb+Pb collisions at $\sqrt{s_{NN}}=2.76$ TeV has been studied in our earlier work \cite{Zhu:2014csa}. In this work, we extend the investigation to non-central collisions in Pb+Pb collisions at both $\sqrt{s_{NN}}=2.76$ and 5.02 TeV in the same framework. We emphasize that there is no parameter to adjust for the intermediate and high $p_T$ regions, since all details on minijets are carried over from Ref. \cite{Zhu:2014csa} and specified in Sec. \ref{RM}. For thermal partons, the inverse slopes $T$ and $T_s$ are independent of centrality \cite{Hwa:2018qss}, so the unknown parameters are just the centrality dependence of normalization factors, $C$ and $C_s$. 

\begin{figure*}[pht]
  \includegraphics[width=0.9\textwidth]{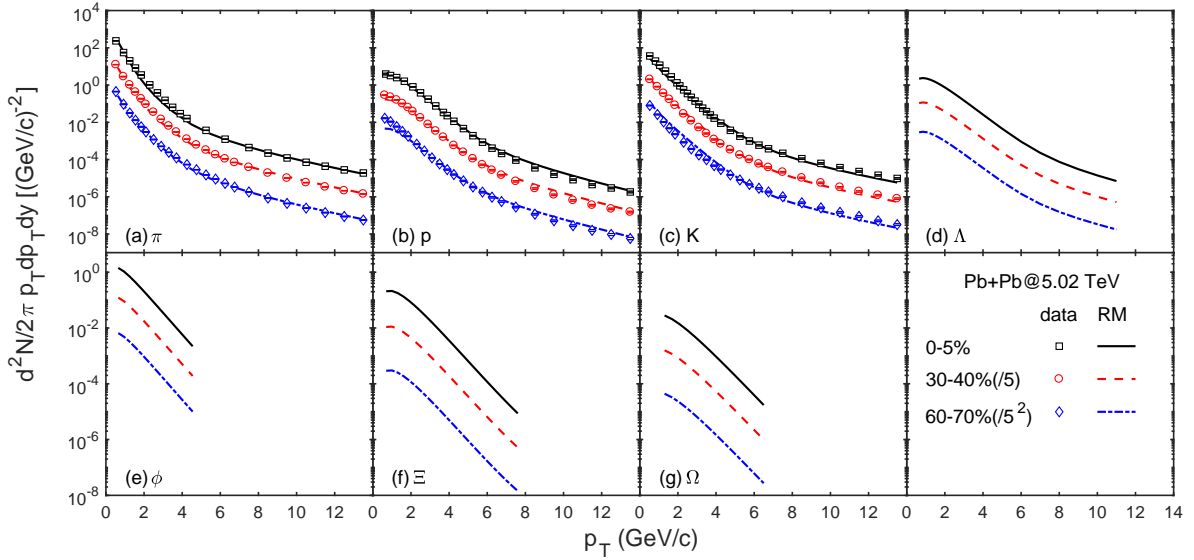}
  \caption{(Color online) Transverse momentum spectra of pion (a), proton (b), kaon (c), $\Lambda$ (d), $\phi$ (e), $\Xi$ (f) and $\Omega$ (g) from the recombination model at midrapidity for Pb+Pb collisions at $\sqrt{s_{NN}}=5.02$ TeV and centralities of 0-5\% (left panel), 30-40\% (middle panel) and 60-70\% (right panel). The experimental data for pion, kaon and proton are taken from Ref. \cite{Acharya:2019yoi}.}
\label{fig6}
\end{figure*}

\begin{table*}
\tabcolsep0.2in
\begin{tabular}{cccc}
\hline
 $\sqrt{s}$ [TeV] & centrality  &  $C$ [(GeV/c)$^{-1}$]& $C_s$  [(GeV/c)$^{-1}$]\\ 
 \hline
  &0-5\%  &23.2 &11.0 \\
2.76&10-20\% &19.5 &9.2 \\
&20-40\%  &15.5 &7.33 \\
&40-60\%  &11.0 &5.04 \\
 \hline
&0-5\%  &22.0 &10.0 \\
 5.02&30-40\% &13.4 &6.55 \\
&60-70\%  &6.8 &3.40 \\
 \hline
 \end{tabular}
 \caption{Parameters $C$ and $C_s$ for Pb+Pb collisions at $\sqrt{s}=2.76$ and 5.02 TeV, respectively.} 
 \label{tab3}
 \end{table*}

The transverse momentum spectra of pion and proton in Pb+Pb collisions at $\sqrt{s_{NN}}=2.76$ TeV and centrality of 10-20\% are shown in Figs. \ref{fig3} and \ref{fig4}, respectively. The data are taken from the ALICE collaboration \cite{Adam:2015kca}. The thermal and shower components in various combinations are shown by different line types. The blue dashed lines show the TT and TTT components, whose magnitudes are determined by fitting the experimental data. What one can see is that the pion distribution without the resonance contribution is lower than the data for $p_T<2$ GeV/c in Fig. \ref{fig3}. Therefore, the extra term $u(p_T, N_{part})$ is inserted in Eq. (\ref{3.2}) for pions,
\begin{eqnarray}
u(p_T, N_{part})=(2.8+0.003N_{part})e^{-p_T/0.45},
\end{eqnarray}
where $N_{part}$ is the number of participants. The number of $0.45$ has the same dimension of momentum and $p_T/0.45$ is dimensionless. The parameters in the above equation are consistent with those used in Ref. \cite{Zhu:2014csa} for central collisions, i.e., $3.95 e^{-p_T/0.45}$. Now, the dependence on $N_{part}$ is introduced.  After taking into account the contribution from resonance decays, the solid line in Fig. \ref{fig3} can reproduce the pion distribution for the whole $p_T$ region very well. The only new adjustable parameter used to fit both the pion and proton spectra is the normalization factor $C$ for the non-strange thermal distribution in Eq. (\ref{2.5}), the value for which is given below. Note how well the proton distribution is  reproduced in Fig. \ref{fig4} with the same thermal and shower parton distributions. It is remarkable that a good agreement with data can be achieved over such a wide range of $p_T$, up to $\sim$14 GeV/c. Comparing the various components of pion and proton for central collisions in Ref. \cite{Zhu:2014csa}, one can see that the results are very similar. The thermal and shower recombination is tremendously important for the hadron production at intermediate $p_T$ region. We can continue along the same line as above to study the strange hadron spectra, i.e., $K$, $\Lambda$, $\phi$, $\Xi$ and $\Omega$. The explicit formulae for the various components of $\Lambda$, $\Xi$, $\Omega$ and $\phi$ are shown in Appendix A-D of Ref. \cite{Zhu:2014csa}.  We will not reproduce them here. The distributions for $K$ production have been given at the end of the previous section. The inverse slope $T_s$ is determined by Eqs. (\ref{2.6})-(\ref{2.8}) and shown in TABLE \ref{tab1}. Hence, $C_s$ is the only one adjustable parameter to fit the data on $K$  (with the spectra of all other strange particles being calculated without adjustments). The values of both $C$ and $C_s$ for the centralities of 0-5\%, 10-20\%, 20-40\% and 40-60\% are given in TABLE \ref{tab3}.

Figure \ref{fig5} shows our results for the transverse momentum spectra of seven identified hadrons, i.e.,$\pi$, $p$, $K$, $\Lambda$, $\phi$, $\Xi$ and $\Omega$, in Pb+Pb collisions at $\sqrt{s_{NN}}=2.76$ TeV and 0-5\%, 10-20\%, 20-40\% and 40-60\% centralities. The distributions for the centrality of 0-5\% are taken from our earlier work \cite{Zhu:2014csa}. Evidently, the agreement with data is excellent for all $p_T$ where data exist \cite{Adam:2015kca, Abelev:2014uua, Abelev:2013xaa, ABELEV:2013zaa} . To our knowledge this is an achievement that has never been reached before.

It is natural to infer from Fig. \ref{fig5} that the recombination model works for non-central collisions in Pb+Pb collisions at $\sqrt{s_{NN}}=2.76$ TeV and should be extended to higher colliding energy $\sqrt{s_{NN}}=5.02$ TeV, where the dynamical effect of energy loss is stronger. Before determining the $\gamma_i$ factor for $\sqrt{s_{NN}}=5.02$ TeV, let us first look at the nuclear modification factor $R_{AA}$ for charged particles in Pb+Pb collisions measured at $\sqrt{s_{NN}}=2.76$ TeV and 5.02 TeV \cite{Acharya:2018}. $R_{AA}$ has a strong centrality dependence and is very similar in magnitude for the two collision energies. In 0-5\% central collisions the yield is suppressed by a factor of about 8 ($R_{AA}\sim$ 0.13) at $p_T=6 \sim 7$ GeV/c. Above $p_T=7$ GeV/c, there is a significant rise of the nuclear modification factor, which reaches a value of about 0.4 around $p_T=40$ GeV/c. Similar behavior can be found for the other centralities. Thus we still choose $q_0=7$ GeV/c for $\sqrt{s_{NN}}=5.02$ TeV. To fit the spectra of pion and proton for intermediate $p_T$ at central collisions, we use $\gamma_0=4.5$. Figure \ref{fig6} shows the transverse momentum spectra of  $\pi$, $p$, $K$, $\Lambda$, $\phi$, $\Xi$ and $\Omega$ in Pb+Pb collisions at $\sqrt{s_{NN}}=5.02$ TeV. The normalization factors $C$ and $C_s$ for various centralities are given in TABLE \ref{tab3}.
 The calculated distributions agree well with the ALICE data for $p_T$ up to 14 GeV/c in Pb+Pb collisions at $\sqrt{s_{NN}}=5.02$ TeV for the centralities of 0-5\%, 30-40\% and 60-70\%. With the retuned parameters $C$ and $C_s$, we show also in Fig. \ref{fig6} our prediction for the spectra of $\Lambda$, $\phi$, $\Xi$ and $\Omega$ in central and non-central Pb+Pb collisions at $\sqrt{s_{NN}}=5.02$ TeV before the experimental data become available.

It is remarkable that the theoretical curves agree with the experimental data in Figs. \ref{fig5} and \ref{fig6} by the adjustment of just two parameters $C$ and $C_s$ in the centrality dependence of the thermal distribution. In each case, the TS, TTS and TSS components play crucial roles in uplifting the spectra in the intermediate region between low $p_T$ where TT and TTT dominate and high $p_T$ where SS and SSS dominate. That aspect of the $p_T$ behavior has become the hallmark of the success of the recombination model, now extended to all centralities. 

A point worthy of special attention is that the values of the inverse slopes obtained in our analysis are high compared to the final temperatures considered in hydrodynamical models, as evidenced by the numbers in TABLE \ref{tab1}. We first note that the relevant ranges of $p_T$ in the two models are different, i.e., $p_T< 2$ GeV/c for HYDRO and $p_T< 4$ GeV/c for TT and TTT in RM without significant contributions from  shower partons, as can be seen in Figs. \ref{fig3} and \ref{fig4}. We next note that for $\Omega$ production, where shower $s$ quarks are rarely produced, the calculated distribution comes entirely from TTT recombination of all $s$ quarks \cite{hwaconf2017}. Thus only thermal partons contribute to $\Omega$ production in RM and the value of the relevant inverse slope is greater than 0.5 GeV/c, valid up to $p_T > 6$ GeV/c, perhaps higher. By considering the widest ranges of $p_T$ in our treatment, we have discovered large values of inverse slopes that should hint at higher temperatures in any study that assumes the medium to be in equilibrium.

\section{summary}\label{summary}
In this paper, we have studied seven identified hadron spectra in relativistic heavy ion collisions at LHC in the recombination model. Results from our calculations are found to describe very well the centrality dependence of $p_T$ spectra of $\pi$, $p$, $K$, $\Lambda$, $\phi$, $\Xi$ and $\Omega$ in Pb+Pb collisions at $\sqrt{s_{NN}}=2.76$ and $\pi$, $p$ and $K$ at $\sqrt{s_{NN}}=5.02$ TeV, respectively. We have also predicted the spectra of $\Lambda$, $\phi$, $\Xi$ and $\Omega$ at various centralities in Pb+Pb collisions $\sqrt{s_{NN}}=5.02$ TeV, which can be compared with experimental measurements in the near future. The geometry and nuclear medium produced in heavy-ion collisions are complex, but the fact that our results from RM agree well with the measured data at all centralities and $p_T$ for the seven identified hadrons gives support to the reliability of the dynamical roles that minijets and their shower partons play in our model.


The fact that the inverse slopes of our thermal distributions are greater than 0.4 GeV/c suggests that when a wide range of $p_T$ is considered in a treatment of equilibrium systems the relevant temperature may be higher than what is obtained in the conventional hydrodynamical models.

The agreement between our model calculations and available data at LHC indicates that the recombination model, together with proper momentum degradation of hard or semihard partons, supports the picture that the centrality dependence of hadron production for the whole $p_T$ region can be described by the recombination of thermal and shower partons in relativistic heavy ion collisions at LHC. In this sense, we can conclude that the recombination model is one of the optimal approaches to describe the hadron production in high energy heavy-ion collisions.

\section*{Acknowledgements}
This work was supported in part by the National Natural Science Foundation of China under Grant No. 11905120.

\end{document}